\documentclass[3p,times,twocolumn
]{elsarticle}

\usepackage{relsize}
\RequirePackage{xspace}

\def\babar{\mbox{\slshape B\kern-0.1em{\smaller A}\kern-0.1em
    B\kern-0.1em{\smaller A\kern-0.2em R}}}
\def\pep2{PEP-II}

\def\kk2f       {\mbox{\tt KK2f}\xspace}
\def\tauola     {\mbox{\tt Tauola}\xspace}
\def\photos     {\mbox{\tt Photos}\xspace}
\def\evtgen     {\mbox{\tt EVTGEN}\xspace}
\def\jetset     {\mbox{\tt JETSET}\xspace}
\newcommand{\Nobs}      {\ensuremath{N_{\rm obs}}}

\newcommand{\thickhline}{%
    \noalign {\ifnum 0=`}\fi \hrule height 1pt
    \futurelet \reserved@a \@xhline
}

\newcommand{\jprlBase}       {Phys.\ Rev.\ Lett.\xspace}
\newcommand{\jprBase}        {Phys.\ Rev.\xspace}
\newcommand{\jplBase}        {Phys.\ Lett.\xspace}
\newcommand{\nimBaseA}       {Nucl.\ Instrum.\ Methods Phys.\ Res., Sect.\ A\xspace}

\newcommand{\nima}      [1]  {\nimBaseA~{\bf #1}}

\newcommand{\cpc}       [1]  {{Comput.\ Phys.\ Commun.\ {\bf #1}}}

\usepackage{amssymb}
\usepackage[figuresright]{rotating}

\begin{document}
\begin{frontmatter}

\title{Measurement of the branching fractions of radiative leptonic $\tau$ decays $\tau \rightarrow l \gamma \nu \bar \nu$, 
($l= e, \mu$) at \babar}

\author[label1,label2]{Benjamin Oberhof\ead{benjamin.oberhof@slac.stanford.edu}}
\author{\\on behalf of the \babar\ collaboration}
\address[label1]{Dipartimento di Fisica Universit\'a di Pisa, Pisa, Italy}
\address[label2]{Laboratori Nazionali dell'INFN, Frascati, Italy}

\begin{abstract}
We perform a measurement of the branching fractions for $\tau \rightarrow l \gamma \nu \bar \nu$, ($l= e, \mu$) decays 
for a minimum photon energy of 10 MeV in the $\tau$ rest frame 
using 430 fb$^{-1}$ of $e^+ e^-$ collisions collected at the center-of-mass energy of the $\Upsilon(4S)$ resonance 
with the \babar\ detector at the \pep2\ storage rings. 
We find $\mathcal B (\tau \rightarrow \mu \gamma \nu \nu) = (3.69 \pm 0.03 \pm 0.10) \times 10^{-3}$ 
and $\mathcal B(\tau \rightarrow e \gamma \nu \nu) = (1.847 \pm 0.015 \pm 0.052) \times 10^{-2}$ 
where the first quoted error is statistical and the second is systematic. 
These results represent a substantial improvement with respect to existing measurements for both channels. 
\end{abstract}


\end{frontmatter}

Leptonic $\tau$ decays are generally well suited to investigate the Lorentz structure of 
electroweak interactions in a model-independent way \cite{michel}. 
Leptonic radiative decays $\tau \rightarrow l \gamma \nu \bar \nu$, ($l= e, \mu$), in particular, 
have been studied for a long 
time \cite{laursen} because they are sensitive 
to the anomalous magnetic moment of the $\tau$ 
\cite{passera}. 
Currently, the branching fraction of the 
$\tau \rightarrow e \gamma \nu \nu$ decay has been measured only by the CLEO collaboration \cite{cleo} 
using 4.68 fb$^{-1}$ of $e^+ e^-$ collisions. 
For a minimum photon energy $E_{\gamma, min}=10$ MeV in the $\tau$ rest frame they 
quote the result $(1.75 \pm 0.06 \pm 0.17) \times 10^{-2}$, where the first error is statistical and the second systematic. 
The CLEO collaboration also made the most precise branching fraction measurement of 
$\tau \rightarrow \mu \gamma \nu \nu$ decay for a minimum photon energy 
in the $\tau$ rest frame $E_{\gamma, min}=10$ MeV, $(3.61 \pm 0.16 \pm 0.35) \times 10^{-3}$. 
In addition, the OPAL collaboration found 
$\mathcal B (\tau \rightarrow \mu \gamma \nu \bar \nu) = (3.0 \pm 0.4 \pm 0.5) \times 10^{-3}$ 
for a minimum photon energy of 20 MeV in the $\tau$ rest frame \cite{opal}. 
In the present work, we perform a measurement of $\tau \rightarrow l \gamma \nu \bar \nu$,  
branching fractions 
for a minimum photon energy of 10 MeV in the $\tau$ rest frame. 

\newpage

This analysis uses data recorded 
by the \babar\ detector at the \pep2\ asymmetric-energy $e^+e^-$ 
storage rings operated at the SLAC National Accelerator Laboratory. 
The data sample consists of 431 fb$^{-1}$ of $e^+ e^-$ collisions recorded at 
$\sqrt{s} = 10.58 $ GeV/$c$. 
The expected cross section for 
$\tau$-pair production is $\sigma_{\tau\tau} = 0.919\pm0.003$ nb 
\cite{tautau} corresponding to a data sample of about 400 million $\tau$-pairs. 
A detailed description of the \babar\ detector is given elsewhere \cite{detector}. 
For this analysis, a Monte Carlo (MC) simulation 
has been used to estimate the signal efficiency and to optimize the search. 
Simulated $\tau$-pair events 
are generated using \kk2f \cite{kk}  
and $\tau$ decays are simulated with \tauola \cite{tauola}. 
Final-state radiative effects in $\tau$ decays are simulated using \photos \cite{photos}. 
A $\tau$-pair MC sample is generated where each $\tau$ lepton decays to 
a mode based on current experimental knowledge \cite{PDG}. 
A separate $\tau$-~pair MC sample is generated where one of the $\tau$ 
leptons decays to $\tau \rightarrow l \gamma \nu \bar \nu$, and the other 
decays according to known decay modes. 
We exclude signal events in the former sample to obtain a $\tau$-pair background sample. 
In this article, charge conjugation is always assumed. 
The MC simulated backgrounds samples include $\mu^+ \mu^-$, $q \bar q$ ($u \bar u$, 
$d \bar d$, $s \bar s$, $c \bar c$), and $B \bar B$ ($B=B^+$, $B^0$) events, where 
$\mu^+ \mu^-$ events are generated by \kk2f \cite{kk}, $q \bar q$ events are generated 
using the \jetset generator \cite{jetset} while $B \bar B$ events are simulated with \evtgen \cite{evtgen}. 
The detector response is simulated with \mbox{\tt GEANT4}~\cite{geant}. 
Two-photon and Bhabha backgrounds are estimated directly from the data. 

The signature for $\tau \rightarrow l \gamma \nu \bar \nu$ decays is a charged 
particle (``track"), identified either as an $e$ or a $\mu$ 
and a neutral deposit in the calorimeter (``cluster") 
such that mass and energy of the lepton-photon pair are compatible with that of the parent $\tau$ lepton. 
Events with two well-reconstructed tracks and 
zero total charge are selected, where no track pair is 
consistent with being a photon conversion in the detector 
material. The polar angle of each track in the laboratory frame is required to be 
within the calorimeter acceptance range to ensure a good particle identification. 
For every track, we require $p_T > 0.3$ GeV/c and a missing transverse 
momentum of $p_{T,miss} > 0.5$ in the event. 
All neutral clusters are required to have a minimum energy of 50 MeV. 
We also reject events with neutral clusters with less than $110$ MeV if they are closer than 25 cm 
from a track, where the distance is measured on the inner wall of the EMC.

Each event is divided into hemispheres 
("signal-" and "tag-" hemisphere) in the center-of-mass (CM) frame by a plane 
perpendicular to the thrust axis, calculated using all reconstructed 
charged and neutral particles \cite{thrust}. 
The signal hemisphere must contain exactly one track and one neutral cluster. 
The tag hemisphere must contain exactly one charged track, identified 
either as an electron, muon or pion, and possibly one additional neutral cluster 
or $n \pi^0$ ($n=$1, 2). Each $\pi^0$ is built up from a pair of neutral clusters 
with invariant mass consistent with that of the $\pi^0$. 
To suppress $e^+ e^- \rightarrow \mu^+ \mu^-$ and Bhabha events, we reject events 
in which the signal-side and tag-side leptons have the same flavor. 
In the signal hemisphere, we require for both channels that the 
distance between the charged track and the neutral cluster 
on the inner wall of the EMC is less than 100 cm. 
For every event, the magnitude of the thrust is required to be within 0.9 and 0.995. 
The lower limit on the thrust magnitude helps to reject most $q \bar q$ events while the upper limit 
helps to reject $e^+ e^- \rightarrow \mu^+ \mu^-$ and Bhabha events. 
For the same reason, we impose the total reconstructed energy to be less than 9 GeV. 

Electrons are identified applying a multivariate algorithm using as input the 
ratio of calorimeter energy to the magnitude of the vector momentum of the track $(E/p)$, the ionization 
loss in the tracking system ($dE/dx$), and the shape of the shower in the calorimeter. 
Muon identification makes use of a bagged decision tree (BDT) 
algorithm \cite{BDT}, which uses as input the number of hits in the IFR, the number of interaction lengths traversed,
and the energy deposition in the calorimeter. Since muons with momenta less than $500$ MeV/$c$ do not 
penetrate into the IFR, the BDT uses also information 
obtained from the inner trackers
to maintain a very low $\pi-\mu$ misidentification probability with high selection efficiencies.
The electron and muon identification efficiencies are 91\% and 62\%, respectively. 
The probability for a $\pi$ to be misidentified as an $e$ in 
$\tau$ decays is below 0.1\%, while the probability to be misidentified 
as a $\mu$ is around 1\% depending on momentum. 

\begin{figure}[htb]
\includegraphics[width=0.50\textwidth]{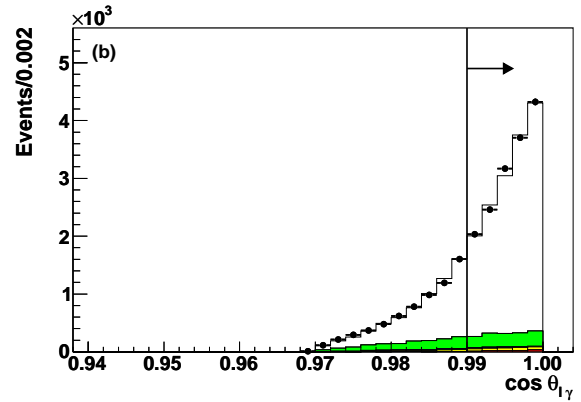}
\caption{
Cosine of the angle between the lepton and photon momenta in the CM frame 
for radiative $\tau$ decay into a muon after applying all selection criteria 
except the one on the plotted quantity. The selection criteria on the plotted quantity 
are highlighted by the vertical lines; we retain the regions indicated by the horizontal arrow. 
In green $\tau \rightarrow \mu \nu \bar \nu$ decays, in blue 
$\tau \rightarrow \pi \pi^0 \nu$ decays, in red $e^+ e^- \rightarrow \mu^+ \mu^-$ events, 
in white signal $\tau \rightarrow \mu \gamma \nu \bar \nu$ decays, and in yellow 
other (background) $\tau$. The black dots are data.}
\label{fig1}
\end{figure}

\begin{figure}[htb]
\includegraphics[width=0.50\textwidth]{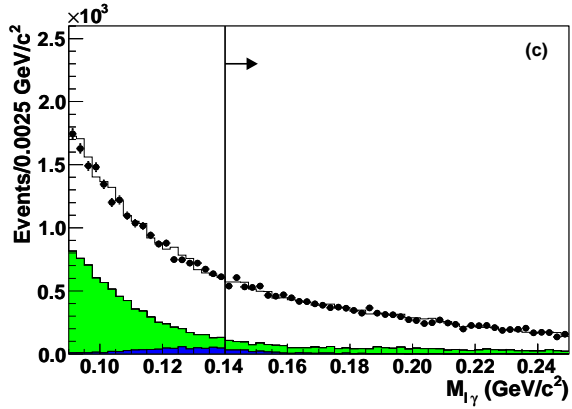}
\caption{
Invariant mass of the lepton photon pair
for radiative $\tau$ decay into an electron after applying all selection criteria 
except the one on the plotted quantity. The selection criteria on the plotted quantity 
are highlighted by the vertical lines; we retain the regions indicated by the horizontal arrow. 
In green $\tau \rightarrow e \nu \bar \nu$ decays, 
in white signal $\tau \rightarrow e \gamma \nu \bar \nu$ decays, and in blue 
other (background) $\tau$. The black dots are data.}
\label{fig2}
\end{figure}

After the preselection, both samples are dominated by background events. 
For the $\tau \rightarrow \mu \gamma \nu \bar \nu$ sample, the main background sources 
are $\tau \rightarrow \mu \nu \bar \nu$, $\tau \rightarrow \pi \pi^0 \nu$ decays, 
$e^+ e^- \rightarrow \mu^+ \mu^-$ events, 
and $\tau \rightarrow \pi \nu$ decays. 
For the $\tau \rightarrow e \gamma \nu \bar \nu$ sample, almost all background contribution is from 
$\tau \rightarrow e \nu \bar \nu$ decays in which the electron radiates a photon 
in the magnetic field of the detector (bremsstrahlung). 
Further background suppression is obtained by placing requirements on the angle between 
the lepton and photon in the CM frame ($\cos \theta_{l \gamma}$). 
For $\tau \rightarrow \mu \gamma \nu \bar \nu$ we require $\cos \theta_{l \gamma} > 0.99$, 
while for $\tau \rightarrow e \gamma \nu \bar \nu$ we require $\cos \theta_{l \gamma} > 0.97$ 
(see Figs. \ref{fig1} and \ref{fig2}). 
To reject background from $\tau \rightarrow e \nu \bar \nu$ decays 
in the $\tau \rightarrow e \gamma \nu \bar \nu$ sample, we further impose 
a minimum value for the invariant mass of the lepton-photon pair $M_{l \gamma} \ge 0.14$~GeV/$c^2$ for this channel. 

In addition to the afore-mentioned quantities, 
the selection criteria in both channels use the energy of the photon and the 
distance between the track and neutral deposit $d_{l \gamma}$ on the 
inner EMC wall. 
The selection criteria are optimized in order to give the smallest 
statistical and systematic uncertainty on the branching fractions. 
After optimization, for $\tau \rightarrow \mu \gamma \nu \bar \nu$, we require 
$\cos \theta_{l \gamma} \ge 0.99$, $ 0.10 \le E_{\gamma} \le 2.5$ GeV, $ 6 \le d_{l \gamma} \le 30$ cm 
and $M_{l \gamma} \le 0.25$ GeV/$c^2$.
For the $\tau \rightarrow e \gamma \nu \bar \nu$ channel, we require 
$\cos \theta_{l \gamma} \ge 0.97$, $ 0.22 \le E_{\gamma} \le 2.0$ GeV, $8 \le d_{l \gamma} \le 65$ cm 
while the lower cut on the invariant mass is set to $M_{l \gamma} \ge 0.14$ GeV/$c^2$.
The selection efficiency determined using the MC samples is given in Table~\ref{tab:results}.

\begin{table}
\begin{center}
\begin{tabular}{lcc}
\hline
Mode & $\tau \rightarrow \mu \gamma \nu \bar \nu$ & $\tau \rightarrow e \gamma \nu \bar \nu$ \\
\hline
Efficiency (\%) &  $0.480 \pm 0.010 $  & $0.105 \pm 0.003 $  \\
$B/N$ &  $0.102 \pm 0.002$  &  $0.156 \pm 0.003$ \\
$N_{exp}$ &  $15649 \pm 125$   &  $18115 \pm 135$  \\
\Nobs &  $ 15688 $ &  $ 18149 $ \\
\hline
\end{tabular}
\caption{Signal efficiency, background contribution $B/N = N_{bkg}/(N_{sig} + N_{bkg})$, 
where $N_{sig}$ are signal and $N_{bkg}$ background events, 
number of expected events ($N_{exp} = N_{sig} + N_{bkg}$) 
and number observed events (\Nobs) for the two decay mode after applying all selection criteria.}
\label{tab:results}
\end{center}
\end{table}

The branching fraction is determined using
\begin{center}
\begin{equation}
\mathcal B_l = \frac{N_{obs} - N_{bkg}}{2 \mbox{ } \sigma_{\tau \tau} \mbox{ } \mathcal L \mbox{ } \epsilon}
\nonumber
\end{equation}
\end{center}
where 
$N_{obs}$ is the number of observed events, $N_{bkg}$ is the number of expected background events, 
$\sigma_{\tau \tau}$ is the cross section for $\tau$ pair production, $\mathcal L$ is the total integrated luminosity 
and $\epsilon$ is the signal efficiency determined from the MC. 
After applying all selection criteria we find 
\begin{equation}
\mathcal B (\tau \rightarrow \mu \gamma \nu \bar \nu) = (3.69 \pm 0.03 \pm 0.10) \times 10^{-3}
\nonumber
\end{equation}
\begin{equation}
\mathcal B (\tau \rightarrow e \gamma \nu \bar \nu) = (1.847 \pm 0.015 \pm 0.052) \times 10^{-2}
\nonumber
\end{equation} 
where the first error is statistical and the second is systematic. 
Efficiency, background expectations ($N_{bkg}$) and the number of observed events (\Nobs)
are shown in Table~\ref{tab:results}. 

Uncertainties on signal efficiency 
estimation and on the number of the expected background events affect the final result. 
For background estimation, we define control regions that are enhanced with background events. 
For $\tau \rightarrow \mu \gamma \nu \bar \nu$, where the major background contribution 
is not peaking in $\cos \theta_{\mu \gamma}$, 
we invert the cut on $\cos \theta_{\mu \gamma}$. 
For $\cos \theta_{\mu \gamma} < 0.8$, 
the maximum expected signal rate is 3\% of the corresponding background rate. 
The maximum discrepancy between the MC sample prediction and the number of observed events is 8\%, 
with an excess of events in the MC sample. 
We take this discrepancy as estimate of the uncertainty on background prediction.
For $\tau \rightarrow e \gamma \nu \bar \nu$ whose major background contributions have similar 
$\cos \theta_{e \gamma}$ distributions as signal, we apply a similar strategy after 
requiring the invariant mass $M_{l \gamma} < 0.14$ GeV/$c^2$; in this case 
we take $\cos \theta_{e \gamma} < 0.90$. The maximum contamination of signal events in this region 
is 10\%, and the maximum discrepancy between the prediction and the number of observed events is 4\% 
with an excess of data events. We take this value as an estimate of the uncertainty on background rate. 
The error on the branching fractions due to the uncertainty on background estimates 
are 0.9\% for $\tau \rightarrow \mu \gamma \nu \bar \nu$, and 0.7\% for $\tau \rightarrow e \gamma \nu \bar \nu$, 
respectively (Table~\ref{tab:syst}). 
Cross-checks of the background estimation are performed by considering the number of events expected and observed in 
different sideband regions immediately neighboring the signal region for each decay mode and found to be compatible with 
the aforementioned systematic uncertainties. 

The most important contributions to the error on efficiency come from 
the uncertainties on particle identification and photon detection efficiency. 
Uncertainties on particle identification efficiency 
are estimated on data control samples, by measuring the variation of the data and MC efficiencies for 
tracks with the same kinematic properties. 
The uncertainty on the efficiency of the electron identification is 
evaluated using a control sample consisting of radiative and non-radiative Bhabha events, while the uncertainty 
for muons is estimated using an $e^+ e^- \rightarrow \mu^+ \mu^- \gamma$ control sample. 
The uncertainty on the pion misidentification probability, as muon 
or electron, is evaluated using samples of $\tau \rightarrow \pi \pi \pi \nu$ decays. 
The corresponding systematic error on the efficiency 
for $\tau \rightarrow l \gamma \nu \bar \nu$ is 1.5\% for both channels. 
To estimate the uncertainty on photon detection efficiency, we rely on two different processes depending on photon energy: 
for high energy photons we use $e^+ e^- \rightarrow \mu^+ \mu^- \gamma$ events while for 
low energy photons we extract the uncertainty from $\pi^0$ reconstruction efficiency. 
in which the photon kinematics can be fully reconstructed using the muon pair. 
Using fully reconstructed $e^+ e^- \rightarrow \mu^+ \mu^- \gamma$ 
events, data and MC are found to be compatible within 1\% for photon energies above 1 GeV. 
For photon energies below 1 GeV we measure 
the $\pi^0$ reconstruction efficiency from 
the ratio of the branching fractions 
for $\tau \rightarrow \pi \nu$ and $\tau \rightarrow \rho \nu$ decays. 
The resulting uncertainty on the $\pi^0$ reconstruction efficiency is found to be below 3\%. 
Including the 1.1\% uncertainty on the branching fractions, 
the resulting uncertainty on single photon detection efficiency is 1.8\%. 
We take this last value as systematic contribution to the efficiency 
for $\tau \rightarrow l \gamma \nu \bar \nu$ due to photon detection efficiency. 
Another possible source of systematic uncertainty arises from the choice of the selection criteria; 
for $\tau \rightarrow e \gamma \nu \bar \nu$ we observe a maximum deviation of 2\% from the 
mean value of the branching fraction depending on the value of $\cos \theta_{l \gamma}$ used for selection. 
We take this value as estimate for the uncertainty on the efficiency due to the choice 
of selection criteria. A similar study on $\tau \rightarrow \mu \gamma \nu \bar \nu$ shows that in this case there is 
no dependence of the result from selection criteria and thus the corresponding uncertainty is negligible. 
All other sources of uncertainty in the signal efficiency are found to be 
smaller than $1.0 \%$, including limited MC 
statistics, track momentum resolution, observables used in 
the selection criteria, and knowledge of the tau branching fractions. 

\begin{table}[t]
\begin{center}
\begin{tabular}{lcc}
\hline
 & $\tau \rightarrow \mu \gamma \nu \bar \nu$ & $\tau \rightarrow e \gamma \nu \bar \nu$\\
\hline
Selection Criteria & -- & 2.0 \\
Photon efficiency & 1.8  & 1.8 \\
Particle Identification & 1.5  & 1.5 \\
Background Evaluation & 0.9  & 0.7 \\
PDG BF & 0.7  & 0.7 \\
$N_{\tau \tau}$ $\tau$ pairs & 0.6  & 0.6 \\
MC Statistics & 0.5  & 0.6 \\
Trigger Selection & 0.5  & 0.6 \\
Track Reconstruction & 0.3  & 0.3 \\
\hline
Total: & 2.8 & 3.4 \\
\hline 
\end{tabular}
\end{center}
\caption{Summary of systematic contributions to the branching fraction (in relative percent) for the two signal channels.}
\label{tab:syst}
\end{table}

In conclusion, we made a measurement of the branching fractions of the radiative leptonic $\tau$ 
decays $\tau \rightarrow e \gamma \nu \bar \nu$ and 
$\tau \rightarrow \mu \gamma \nu \bar \nu$ 
for a minimum photon energy of 10 MeV in the $\tau$ rest frame 
using the full dataset of $e^+ e^-$ collisions collected 
by \babar\ at the center-of-mass energy of the $\Upsilon(4S)$ resonance. 
We find $\mathcal B (\tau \rightarrow \mu \gamma \nu \nu) = (3.69 \pm 0.03 \pm 0.10) \times 10^{-3}$ 
and $\mathcal B(\tau \rightarrow e \gamma \nu \nu) = (1.847 \pm 0.015 \pm 0.052) \times 10^{-2}$ 
where the first error is statistical and the second is systematic. 
These results represent an improvement of about a factor of three for both channels with respect 
to the previous experimental bounds \cite{cleo}, reducing both statistic and systematic contributions 
for both channels. 
The main contribution to the total error for both measurements 
comes from the photon detection efficiency and particle identification. 
For $\tau \rightarrow e \gamma \nu \bar \nu$, there is also an important contribution 
coming from the dependence of the final result from the selection criteria on the lower value 
of the outgoing electron and the photon. 
Our results are in agreement with the SM values, 
$\mathcal B (\tau \rightarrow \mu \gamma \nu \nu) = (3.686 \pm 0.009) \times 10^{-3}$ 
and $\mathcal B(\tau \rightarrow e \gamma \nu \nu) = (1.843 \pm 0.002) \times 10^{-2}$,  
obtained from the TAUOLA \cite{tauola} MC, 
which uses the PHOTOS \cite{photos} package to simulate QED radiative corrections in the decay up 
to third order in the fine structure constant $\alpha$. 

\bibliographystyle{elsarticle-num}

\begin{thebibliography}{00}
\bibitem{michel} L.~Michel, Proc. Roy. Soc. Lond. A {\bf 63}, 514 (1950).
\bibitem{laursen} M.~L.~Laursen, M.~A.~Samuel, and A.~Sen, \jprBase D {\bf 29}, 2652 (1984). 
\bibitem{passera} M.~Fael {\it et al.}, arXiv:1301.5302, (2013). 
\bibitem{cleo} T.~Bergfeld {\it et al.} (CLEO collaboration), {\jprlBase} {\bf 84}, 830-834 (2000).
\bibitem{opal} G.~Alexander {\it et al.} (OPAL collaboration), \jplBase B {\bf 388}, 437 (1996).
\bibitem{tautau} S.~Banerjee {\it et al.}, \jprBase D {\bf 77}, 054012 (2008).
\bibitem{detector} B.~Aubert {\it et al.} (\babar\ collaboration), Nucl. Instrum. Methods Phys. Res., Sect. {\bf A 479}, 1 (2002).
\bibitem{det2} B.~Aubert {\it et al.} (\babar\ collaboration), Nucl. Instrum. Methods Phys. Res., Sect. {\bf A 729}, 615-701 (2013).
\bibitem{kk} S.~Jadach, B.~F.~Ward, and Z.~Was, \cpc {\bf 130}, 260 (2000).
\bibitem{tauola} S.~Jadach {\it et al.}, \cpc {\bf 76}, 361 (1993).
\bibitem{photos} E.~Barberio and Z.~Was, \cpc {\bf 79}, 291 (1994).
\bibitem{PDG} K.A. Olive {\it et al.} (Particle Data Group), Chin. Phys. C, {\bf 38}, 090001 (2014)
\bibitem{jetset} T.~Sjostrand, S.~Mrenna, and P.~Skands, JHEP, 0605-026 (2006).
\bibitem{evtgen} D.J.~Lange, \nima {\bf 462}, 152-155 (2001).
\bibitem{geant} S.~Agostinelli {\it et al.}, \nima {506}, 250 (2003).
\bibitem{thrust} E.~Farhi, Phys. Rev. Lett. {\bf 39}, 1587 (1977).
\bibitem{BDT} E.~L.~Allwein, R.~E.~Schapire, and Y.~Singer, J.\ Machine Learning Res. {\bf 1}, 113 (2000).
\end{thebibliography}

\end{document}